\documentclass[12pt,tightenlines,floats,aps,prd,twoside,onecolumn,english,
	superscriptaddress,showkeys,nofootinbib,preprintnumbers,showpacs]{revtex4}

\usepackage{babel}
\usepackage{amsmath}
\usepackage{amsfonts}
\usepackage{amssymb}
\usepackage{dsfont}

\newcommand{\be}{\begin{equation}}
\newcommand{\ee}{\end{equation}}

\def\p{\partial}

\def\b{\beta}
\def\g{\gamma}

\def\om{\omega}

\begin{document}

\title{Modifications in the Spectrum of Primordial Gravitational Waves Induced by Instantonic Fluctuations}

\author{\firstname{Simone} \surname{Mercuri}}
\email{simone.m2277@gmail.com}
\affiliation{Institute for Gravitation and the Cosmos, The Pennsylvania State University,\\ 104 Davey Lab, Physics Department, University Park, PA 16802, USA}

\begin{abstract}
Vacuum to vacuum instantonic transitions modify the power spectrum of primordial gravitational waves. We evaluate the new form of the power spectrum for ordinary gravity as well as the parity violation induced in the spectrum by a modification of General Relativity known as Holst term and we outline the possible experimental consequences.
\end{abstract}



\maketitle

\section{General Remarks}\label{Intro}
The great enhancement in the sensitivity of the Cosmic Microwave Background (CMB) polarization detectors expected in the near future could be the key to efficiently probe the existence of a background of primordial gravitational waves (GW).
The two basic modes of the CMB spectrum are generally referred as parity even E-modes, produced by scalar and tensor fluctuations and parity odd B-modes, produced only by tensor fluctuations. So, it is in the B-modes of CMB that we can find information about the existence of primordial GW; needless to say that the predicted amplitude of the produced signal is very small, making its detection an extremely hard task.

Primordial GW are produced by quantum fluctuations around the classical expanding background characterizing the inflationary era. From an experimental perspective, they may be observed by measuring the B-modes self-correlation in the CMB spectrum (BB-correlation). In general, the correlation functions between CMB modes and temperature fluctuations depend on the GW power spectrum, which can be calculated in the canonical quantization scheme by evaluating the variance of the gravitational perturbations with respect to a unique vacuum state (see \cite{Dod03}). Nevertheless, the non-abelian structure of the large component of the local gauge group of gravity suggests that, in general, a class of inequivalent vacua characterized by different winding numbers exists: this fact compels us to reconsider the calculation of the primordial GW power spectrum, which in fact results modified by the existence of instantonic transitions.

Specifically, in this Letter, by using semiclassical arguments based on the Wentzel-Kramer-Brillouin (WKB) approximation, we demonstrate that the vacuum-to-vacuum transitions can produce a modification of the primordial GW power spectrum and, when topological extensions of the Hilbert--Palatini action are taken into account, they generate an effective violation of parity. The violation of parity drastically reflects on the phenomenological outcomes of this model, which predicts non-vanishing TB and EB cross correlations.\footnote{The TB and EB correlations are in fact the only two cross-correlation functions among all the possible combinations of E and B modes and temperature fluctuations, T, that vanish identically as a consequence of classical symmetries, specifically the parity invariance of classical General Relativity.} In general, since the magnitude of the temperature fluctuations, T, is much larger than either E or B fluctuations, it would be easier to find evidence for TB than EB cross-correlation. However, according to our results, since TB as well as EB cross-correlations are related to instantonic transitions, their amplitude is exponentially suppressed with respect to that associated to the other cross-correlations. Therefore, their detection may be a very hard task unless gravity is almost completely chiral, as we detail in what follows.

The Holst action for General Relativity \cite{Hol96} will serve as our prototype to study this mechanism, being aware that other topological modifications of ordinary gravity can produce the same effect.\footnote{Further considerations about parity violating instantonic effects produced by other topological terms will appear in a forth-coming paper, where we will elucidate the role of Nieh--Yan and Chern--Pontryagin terms in this picture.} Nonetheless, the Holst action can be considered particularly interesting as it is the Lagrangian counterpart of the Ashtekar--Barbero reformulation of canonical General Relativity \cite{Ash86-87,Bar95}, which is at the basis of the non-perturbative approaches to the quantization of gravity belonging, let's say, to the Loop Quantum Gravity family \cite{AshLew04,Rov04,Thi06,Mer10}. Therefore, a result indicating that parity can be violated by quantizing the Holst action in the WKB approximation, would suggest that the same effect could characterize the full quantum theory.

Without giving further details, we only mention that the Holst action consists of the standard Hilbert--Palatini action plus an on-(half)shell vanishing term, called Holst modification, which contains a new constant, the Barbero--Immirzi (BI) parameter. Many different interpretations have been proposed for the BI parameter. From our perspective, the most interesting one interprets the BI parameter as an instanton angle \cite{GamObrPul99,Mer06,Mer08,Kau08,Mer09p,DatKauSen09,Sen10,MerRan10}. In these terms, we expect that it enters in the coefficients characterizing the parity violation effects as a chiral parameter, in accordance with the assumption made by Contaldi, Magueijo and Smolin \cite{ConMagSmo08} -- though, as we will show, the modification to the GW power spectrum produced by the Holst modification through instantonic effects is not simply the one assumed in \cite{ConMagSmo08}. 

\section{Instantonic Transitions Amplitude}
Roughly speaking, all the inflationary models are characterized by a ``slow roll phase'' during which the Universe goes through a rapid expansion. The inflaton potential during the slow roll is well approximated by a constant, so that a simple dynamical model that efficiently describes this phase can be derived from the Hilbert--Palatini action for gravity with the cosmological constant, $\Lambda$. Primordial GW are produced during the de Sitter expansion phase by quantum fluctuations around the classical background, namely they are quadratic deviations from the unique vacuum state, $\left|0\right>$. Specifically, by using a harmonic oscillator analogy and following Ref. \cite{Dod03}, we can write the variance of gravitational perturbations as 
\begin{align}\nonumber\label{VGW}
\left<0\right|\hat{h}_{L/R}^{\dagger}(\underline{k},t)\hat{h}_{L/R}(\underline{k}^{\prime},t)\left|0\right>= &(2\pi)^3\frac{16\pi G}{a^2}\left|v_{L/R}(\underline{k},t)\right|^2\delta(\underline{k},\underline{k}^{\prime})
\\
\equiv &(2\pi)^3P^{L/R}_{h}(k)\delta(\underline{k},\underline{k}^{\prime})\,,
\end{align}
where $\hat{h}_{L/R}$ is the operator associated to the left and right modes of the graviton field respectively, $a$ is the conformal factor and $v_{L/R}(\underline{k},t)$ is the expansion coefficient of the quantum operators, namely $\hat{h}_{L/R}(\underline{k},t)=v_{L/R}(\underline{k},t)\hat{a}_{\underline{k}}+v_{L/R}^*(\underline{k},t)\hat{a}^{\dagger}_{\underline{k}}$. The second line of Eq. (\ref{VGW}) defines the GW power spectrum $P_{h}^{L/R}(k)$ for left and right modes, that in the de Sitter space-time reads $P^{L}_h(k)=P^{R}_h(k)=P_h(k)$, where \cite{Dod03}:
\begin{equation}\label{UPS}
P_h(k)=\frac{8\pi G\hbar}{k^3}H^2=\frac{8\pi G\hbar}{3k^3}\Lambda\,.
\end{equation}
Namely, the left and right GW modes turn out to have the same power spectrum.

A closer look at the global gauge structure of the de Sitter solution is now in order. To be specific, we refer here to the spatially flat de Sitter geometry. The de Sitter space-time is characterized by the following spatial boundary conditions
\be\label{BC}
\,^{(3)}\!\!\!\left.R_{\frac{}{}}\!\right|_{\p M}=0\,,
\ee
so that the spatial connection reduces on the boundary to a pure gauge:
\be\label{BC1}
\,^{(3)}\!\!\left.\om_{\frac{}{}}\!\right|_{\p M}=\Omega^{-1} d\Omega\,,
\ee 
and has topology $M=\mathbb{R}\times S^3$; namely, every spatial slice is flat and, through the one-point compactification, topologically equivalent to $S^3$. As a consequence $\p M=S^3_i\cup S^3_f$; $S^3_{i}$ and $S^3_{f}$ being, respectively, the two Cauchy slices where the initial and final states of the system are defined. The vacuum state of such a gravitational system corresponds to pure gauge field configurations of the $SU(2)$ group, which is the natural realization of the topological sphere symmetries plus time reversal.
Considering that $\Pi_3\left[SU(2)\right]=\mathbb{Z}$, we expect that a class of inequivalent vacua, classified according to the corresponding homotopy classes, arise. The homotopy classes are determined by the winding (or instantonic) number through the Maurer--Cartan integral,
\be\label{MCI}
w\left(\Omega\right)=\frac{1}{24\pi^2}\int\limits_{S^3}{\rm Tr}\,\Omega^{-1}d\Omega\wedge\Omega^{-1}d\Omega\wedge\Omega^{-1}d\Omega\,.
\ee
So that, to refer to the full class of inequivalent vacua, we can use the notation $\left|\Omega_w\right>$, $w=0,1,2,\dots$. It is clear that the vacuum states $\left|\Omega_w\right>$ are not fully gauge invariant. In this regard, it is worth noting that $\mathcal{G}_n\left|\Omega_w\right>=\left|\Omega_{w+n}\right>$, where $\mathcal{G}_n$ is a large gauge transformation with winding number $n$. Therefore, as physical vacuum, we consider a linear superposition of the w-vacuum states $\left|\Omega_w\right>$, indicating it as $\left|\Omega\right>$. We note that $\left<\Omega\right|\left.\!\Omega\right>\neq 1$, since different vacua are quantum mechanically bridged by left and right instantons 0-modes: this generates a non-trivial modification in the GW power spectrum as we are going to argue.

By using again the harmonic oscillator analogy, we can generalize the definition in Eq. (\ref{VGW}) to the existence of inequivalent vacua as follows:
\begin{align}\nonumber\label{VWWGW}
\left<\Omega\right|\,&\hat{h}_{L/R}^{\dagger}(\underline{k},t)\hat{h}_{L/R}(\underline{k}^{\prime},t)\!\left|\,\Omega\right>
\\\nonumber
&=(2\pi)^3\frac{16\pi G}{a^2}A\left[\Omega\stackrel{L/R}{\longrightarrow}\Omega\right]\left|v_{L/R}(\underline{k},t)\right|^2\delta(\underline{k},\underline{k}^{\prime})
\\
&\equiv(2\pi)^3\mathcal{P}^{L/R}_{h}(k)\delta(\underline{k},\underline{k}^{\prime})\,,
\end{align}
where $A\left[\Omega\stackrel{L/R}{\longrightarrow}\Omega\right]$ denotes the tunneling amplitude for instantonic transitions associated respectively to left and right 0-modes. So that the generalized power spectrum for primordial gravitational waves reads: 
\begin{equation}\label{MPGWPS}
\mathcal{P}^{L/R}_{h}(k)=A\left[\Omega\stackrel{L/R}{\longrightarrow}\Omega\right]P^{L/R}_{h}(k)\,.
\ee
In accordance with a procedure used by 't Hooft \cite{Hoo76}, the amplitudes of the vacuum-to-vacuum instantonic transitions can be evaluated as tunneling processes in the WKB approximation, i.e., by using the following formula:
\be\label{SCA}
A\left[\Omega\stackrel{L/R}{\longrightarrow}\Omega\right]\approx\sum_{\Delta w} e^{-\left|S^{L/R}_{\rm Eff}\left(\Delta w\right)\right|/\hbar}\,,
\ee
where $\Delta w$ is the absolute variation in the winding number associated to the initial and final vacuum states. $S_{\rm Eff}$ is the Hamilton function associated to the left and right quantum fluctuations in the classically prohibited region, or, equivalently, the pull-back of the self-dual and antiself-dual Euclidean actions to the solution of the equations of motion. The result obtained by 't Hooft indicates that the likelihood of the occurrence of instantonic transitions exponentially decays as the relative Pontryagin number of the different vacua involved in the process increases. On general grounds, the same procedure can be applied to gravity and, in particular, to the Holst modified action for GR, provided that particular care is used in treating the boundary terms. In fact, managing the boundary terms in the second order formulation of GR is far from being trivial \cite{ManMar06} due to the difficulties in reabsorbing subtle infinities \cite{LikSlo09}: only very recently has it been shown that the Palatini first order formulation of gravity, instead, is not characterized by the same limitations \cite{AshEngSlo08}, so that it appears to be a more promising approach to formulating a well-posed problem. Therefore, considering the Palatini formulation of GR with the Holst modification and the cosmological constant, we now evaluate the amplitude of instantonic transitions around de Sitter classical background.

For this purpose, let us start defining the classical Euclidean action. We consider a 4-dimensional Riemannian manifold $M$ with a boundary $\p M$. Let $e=e^{a}_{\mu}\g_a dx^{\mu}$ be the Clifford-valued local basis 1-form, where $\g^a$ indicates the four Dirac gamma matrices. We indicate with $\om=\om_{\mu}^{ab}\Sigma_{ab}dx^{\mu}$ the connection 1-form, where $\om_{\mu}^{ab}$ is the usual spin-connection and $\Sigma^{ab}=\frac{1}{2}\left[\g^a,\g^b\right]$ are the generators of the $SO(4)$ group. The matrices $\Sigma^{ab}$ satisfy the following relations:
\begin{subequations}
\begin{align}
\left[\g^a,\Sigma^{bc}\right]&=4\delta^{a[b}\g^{c]}\,,
\\
\left[\Sigma^{ab},\Sigma^{cd}\right]&=4\left(\delta^{a[c}\Sigma^{d]b}-\delta^{b[c}\Sigma^{d]a}\right)\,.
\end{align}
\end{subequations}
The curvature 2-form associated to the connection is defined as $R=d\om+\frac{1}{4}\,\om\wedge\om$. Hence, remembering that in Euclidean space
\be
{\rm Tr}\gamma^a\gamma^b\gamma^c\gamma^d=4\left(\delta^{ac}\delta^{bd}-\delta^{ad}\delta^{bc}-\delta^{ab}\delta^{cd}\right)\,,
\ee
the Euclidean Hilbert--Palatini action with the Holst modification can be written as (we set $8\pi G=1$; we will reintroduce the correct dimensions when required by the argument):
\begin{align}\label{HAWCC}\nonumber
S\left[e,\om\right]=\,&\frac{1}{16}\int\limits_M {\rm Tr}\,e\wedge e\wedge\left[\star W+\frac{1}{\b}\,W\right]
\\
&-\frac{1}{16}\int\limits_{\p M}{\rm Tr}\,e\wedge e\wedge\left(\star\,\om+\frac{1}{\b}\,\om\right)\,,
\end{align}
where $W=R-\frac{\Lambda}{3!}\,e\wedge e$, $\b$ is the BI parameter and $\Lambda$ is the cosmological constant. In accordance with \cite{AshEngSlo08}, we added a suitable boundary term to make the action finite. By varying the above action with respect to the connection $\om$ and the gravitational field 1-form $e$, we respectively obtain the following equations of motion:
\begin{subequations}
\begin{align}\label{OIICSE}
d\left(e\wedge e\right)+\frac{1}{2}\,\left[\om, e\wedge e\right]&=0\,,
\\\label{EE}
\left[e,\star R\right]-\frac{\Lambda}{3}\,\left[e,\star\left(e\wedge e\right)\right]+\frac{1}{\b}\,\left[e,R\right]&=0\,,
\end{align}
\end{subequations}
where the symbol $\left[\,\cdot\,,\,\cdot\,\right]$ stands for the commutator.\footnote{It is worth noting that $\left[e,e\wedge e\right]=0$.} Eq. (\ref{OIICSE}) leads to the so-called second Cartan structure equation, i.e. 
\be\label{IICSE}
de+\frac{1}{4}\,\left[\om,e\right]=0\,,
\ee
which relates the gravitational field to the spin connection. The curvature associated with the connection satisfying Eq. (\ref{IICSE}) fulfills the following identity:
\be\label{BI}
\left[R,e\right]=0\,.
\ee
Finally, once Eq. (\ref{IICSE}) is satisfied, the remaining equations of motion (\ref{EE}) reduce to the ordinary Euclidean Einstein equation of GR, since the term proportional to the BI parameter vanishes by virtue of Eq. (\ref{BI}). So, as mentioned before, the Holst action is classically equivalent to the Hilbert--Palatini action \cite{CorWil-Ewi10} (at least in pure gravity \cite{Mer06}); nevertheless, this equivalence does not survive to the quantization, even in the WKB approximation as we are going to demonstrate.

Remarkably, the equations of motion (\ref{OIICSE}) and (\ref{EE}) are identically satisfied if the following condition holds:
\be\label{DSC}
R=\frac{\Lambda}{3}\,e\wedge e\,.
\ee
The condition above in fact characterizes a family of solutions of the Euclidean equations of motion with a cosmological constant and can be used to evaluate the tunneling amplitude in Eq. (\ref{SCA}) according to the WKB procedure.

First, let us evaluate the pull-back of the action (\ref{HAWCC}) on the solution of the equations of motion. We have:
\be\label{DSA}
S_{\rm Eff}=\frac{3}{32\cdot 8\pi G\Lambda}\int\limits_{M}{\rm Tr}\left(R\wedge\star R+\frac{1}{\b}\,R\wedge R\right)\,,
\ee
where we reintroduced the physical constants. The presence of the Euler and Pontryagin classes in the expression above already reveals that the quantum tunneling between inequivalent vacua violates parity: in fact, while for purely self-dual modes the two topological terms are equivalent, for the anti-self-dual ones they are opposite in sign. Therefore, the left and right quantum transitions feature a different amplitude. To make this statement explicit, let us rewrite the effective action above extracting the self and anti-self dual contributions. To this purpose we define:
\begin{subequations}
\begin{align}
R^{(+)}=\star R+ R\,,
\\
R^{(-)}=\star R- R\,,
\end{align}
\end{subequations}
which are respectively the self-dual and anti-self-dual components of the curvature 2-form. By using these definitions, we have
\be\label{DSAD}\nonumber
S_{\rm Eff}^{L/R}=\frac{3}{128\cdot 8\pi G\Lambda}\frac{1\pm\b}{\b}\int\limits_{M}{\rm Tr}\left[R^{(\pm)}\wedge R^{(\pm)}\right]\,.
\ee
Therefore, by calculating the integral of the Pontryagin class, we can finally write the amplitude (\ref{SCA}) as
\be\label{SCALR}
A\left[\Omega\stackrel{L/R}{\longrightarrow}\Omega\right]=\sum_{\Delta w\geq 0}\exp\left\{-\frac{9}{128}\left|\frac{\b\pm 1}{\b}\right|\frac{\pi}{\Lambda\hbar G}\Delta w\right\}\,.
\end{equation}
where the $\pm$ signs refer respectively to the left and right 0-modes. It is worth noting that the chiral parameter role is here played by the BI parameter, $\beta$, which, in fact, behaves like an instanton angle (see also \cite{MerRan10}). 

\section{GW Power Spectrum}
We now have all the elements to write the final expression for the primordial GW power spectrum as defined in Eq. (\ref{MPGWPS}). Specifically we obtain:
\begin{equation}
\mathcal{P}^{L/R}_h(k)=P_h(k)\sum_{\Delta w\geq 0}e^{-\frac{9}{128}\left|\frac{\b\pm 1}{\b}\right|\frac{\pi}{\Lambda\hbar G}\Delta w}\,.
\end{equation}
Some comments are now in order. First, let us consider the limit $\b\to\infty$, which corresponds to ordinary General Relativity without the Holst modification. In this case the new GW power spectrum reduces to
\begin{equation}\label{LRSE}
\mathcal{P}^{L/R}_h(k)=\mathcal{P}_h(k)=P_h(k)\sum_{\Delta w\geq 0}e^{-\frac{9}{128}\frac{\pi}{\Lambda\hbar G}\Delta w}\,,
\end{equation}
which, once compared with the one in Eq. (\ref{UPS}), reveals an exponentially decaying modification (due to the presence of instantonic transitions) that depends on the inflation potential, here related to the constant $\Lambda$. As expected, as soon as we neglect instanton vacuum-to-vacuum transitions, namely when $\Delta w=0$, the above expressions reduce to the ordinary one. 

Furthermore, let us consider the two combinations $\mathcal{P}^+_h(k)=\mathcal{P}_h^R(k)+\mathcal{P}_h^L(k)$ and $\mathcal{P}^-_h(k)=\mathcal{P}_h^R(k)-\mathcal{P}_h^L(k)$ which enter in the cross-correlation functions of CMB and read
\begin{subequations}
\begin{align}\nonumber
\mathcal{P}_h^{+}(k)=P_h(k)\sum_{\Delta w\geq 0}&\left(e^{-\frac{9}{128}\left|\frac{\b-1}{\b}\right|\frac{\pi}{\Lambda\hbar G}\Delta w}\right.
\\
&+\left.e^{-\frac{9}{128}\left|\frac{\b+ 1}{\b}\right|\frac{\pi}{\Lambda\hbar G}\Delta w}\right)\,,
\\\nonumber
\mathcal{P}_h^{-}(k)=P_h(k)\sum_{\Delta w\geq 0}&\left(e^{-\frac{9}{128}\left|\frac{\b- 1}{\b}\right|\frac{\pi}{\Lambda\hbar G}\Delta w}\right.
\\
&-\left.e^{-\frac{9}{128}\left|\frac{\b+ 1}{\b}\right|\frac{\pi}{\Lambda\hbar G}\Delta w}\right)\,.
\end{align}
\end{subequations}
Interestingly enough, $\mathcal{P}_h^{-}(k)$ characterizes the TB and EB cross correlations and is in general non-vanishing, thus revealing the possible existence of a parity violation in gravity due to the presence of the Holst modification; let us look at this in detail.

Again, as soon as we set $\Delta w=0$, the above expressions reduce to the ordinary ones. Naturally, the processes with $\Delta w=0$ do not contribute to $\mathcal{P}^-_h(k)$; meaning that no parity violation is expected when instantonic transitions are neglected. The function $\mathcal{P}^-_h(k)$ vanishes, as well, in the case $\Delta w\neq 0$ and $\beta\to\infty$, that describes the effects of instantonic transitions in ordinary gravity as already said above (see Eq. (\ref{LRSE})); specifically for $\b\to\infty$ and $\Delta w \neq 0$ we have
\begin{subequations}
\begin{align}
\mathcal{P}_h^{+}(k)&=2P_h(k)\left(1+\sum_{\Delta w\geq 1}e^{-\frac{9}{128}\frac{\pi}{\Lambda\hbar G}\Delta w}\right)\,,
\\
\mathcal{P}_h^{-}(k)&=0\,.
\end{align}
\end{subequations}
Finally, the parity violating function $\mathcal{P}_h^{-}(k)$ does not vanish only if $\Delta w\neq 0$ and $\left|\beta\right|<+\infty$. Therefore, in general, non-vanishing TB and EB cross-correlations are exponentially suppressed with respect to the other correlation functions, thus making their observation an extremely hard task.\footnote{For the sake of clarity, we stress that this is in contrast with what stated in \cite{ConMagSmo08}, in fact, according to this calculation, there is no reason to support the idea that TB can be measured more easily than BB, which in \cite{ConMagSmo08} is a mere consequence of the specific toy model they used.} One remarkable case is $\left|\beta\right|\approx 1$, namely when gravity is (almost) perfectly chiral. In this case, in fact, we approximately have $\mathcal{P}^+_h(k)\approx P_h(k)(1+\mathcal{P}^-_h(k)/P_h(k))$ or, more interestingly,
\be
D_{L/R}\equiv\frac{\mathcal{P}^-_h(k)}{\mathcal{P}^+_h(k)}=\frac{\sum_{\Delta w\geq 1}e^{-a\Delta w}}{1+\sum_{\Delta w\geq 1}e^{-a\Delta w}}\approx e^{-a}\,,
\ee
where $a=9\pi\left|\beta-1\right|/128\beta\Lambda\hbar G$. In particular, requiring $0.1\leq D_{L/R}\leq 1$, we find a limit on the BI parameter, i.e. $1-\frac{128}{9\pi}\Lambda\hbar G\ln10\leq\beta\leq1+\frac{128}{9\pi}\Lambda\hbar G\ln10$, which gives an indication about the possible observability of the TB cross correlation, considering that $\Lambda\hbar G\propto H^2/M^2_{Pl}\propto\rho/M^4_{Pl}$ is the energy density at the time of inflation in units of the Planck mass.

While this paper was in the referral process, two other papers, \cite{MagBen11} and \cite{BetMag11}, discussed possible parity violation effects in gravity. Even though the effects described in all these works are analogous, their relationship is far from being obvious and worth of being clarified.

\acknowledgments The author would like to thank Joao Magueijo for an interesting and stimulating email exchange we recently had about parity violation in gravity. The author is also grateful to Massimiliano Lattanzi for many interesting discussions about Early Cosmology and particularly indebted to Dana Levine for her precious encouragement and valuable suggestions during the writing of this paper. Finally, the author would like his lack of financial support be known.


\begin{thebibliography}{99}

\bibitem{Dod03}
S. Dodelson, \emph{Modern Cosmology}, Elsevier Academic Press, Oxford, 2003.

\bibitem{Hol96}
S. Holst, Phys. Rev. D \textbf{53}, 5966 (1996).

\bibitem{Ash86-87}
A. Ashtekar, Phys. Rev. Lett. \textbf{57}, 2244 (1986); Phys. Rev. D \textbf{36}, 1587 (1987).

\bibitem{Bar95}
F. Barbero, Phys. Rev. D \textbf{51}, 5498 (1995); Phys. Rev. D \textbf{51}, 5507 (1995).

\bibitem{AshLew04}
A. Ashtekar and J. Lewandowski, Class. Quant. Grav. \textbf{21}, R53 (2004).

\bibitem{Rov04}
C. Rovelli, \emph{Quantum Gravity}, Cambridge University Press, Cambridge, 2004.

\bibitem{Thi06}
T. Thiemann, \emph{Modern Canonical Quantum General Relativity}, Cambridge University Press, Cambridge, 2006.

\bibitem{Mer10}
S. Mercuri, proceeding of the 5th International School on Field Theory and Gravitation, April 20-24, 2009 Cuiab\'a, Brazil, PoS(ISFTG), \textbf{016} (2009).

\bibitem{GamObrPul99}
R. Gambini, O. Obregon, and J. Pullin, Phys. Rev. D \textbf{59}, 047505 (1999).

\bibitem{Mer06}
S. Mercuri, Phys. Rev. D \textbf{73}, 084016 (2006).

\bibitem{Mer08}
S. Mercuri, Phys. Rev. D \textbf{77}, 024036 (2008).

\bibitem{Kau08}
R.K. Kaul, Phys. Rev. D \textbf{77}, 045030 (2008).

\bibitem{Mer09p}
S. Mercuri, arXiv:0903.2270.

\bibitem{DatKauSen09}
G. Date, R.K. Kaul, and S. Sengupta, Phys. Rev. D \textbf{79}, 041901 (2009). 

\bibitem{Sen10}
S. Sengupta, Class. Quantum Grav. \textbf{27}, 145008 (2010).

\bibitem{MerRan10}
S. Mercuri and A. Randono, Class. Quant. Grav. \textbf{28}, 025001 (2011).

\bibitem{ConMagSmo08}
C. R. Contaldi, J. Magueijo, and L. Smolin, Phys. Rev. Lett. \textbf{101}, 141101 (2008).

\bibitem{Hoo76}
G. 't Hooft, Phys. Rev. D \textbf{14}, 3432 (1976).

\bibitem{ManMar06}
R. B. Mann and D. Marolf, Class. Quant. Grav. \textbf{23}, 2927 (2006).

\bibitem{LikSlo09}
T. Liko and D. Sloan, Class. Quant. Grav. \textbf{26}, 145004 (2008).

\bibitem{AshEngSlo08}
A. Ashtekar, J. Engle, and D. Sloan, Class. Quant. Grav. \textbf{25}, 095020 (2008).

\bibitem{CorWil-Ewi10}
A. Corichi and E. Wilson-Ewing, Class. Quant. Grav. \textbf{27}, 205015 (2010).

\bibitem{MagBen11}
J. Magueijo and D. M. T. Benincasa, Phys. Rev. Lett. \textbf{106}, 121302 (2011).

\bibitem{BetMag11}
L. Bethke and J. Magueijo, arXiv:1104.1800.

\end{thebibliography}
\end{document}